\input phyzzx
\overfullrule=0pt
\hsize=6.5truein
\vsize=9.0truein
\voffset=-0.1truein
\hoffset=-0.1truein
%
%

\def\IC{{\ \hbox{{\rm I}\kern-.6em\hbox{\bf C}}}}
\def\IR{{\hbox{{\rm I}\kern-.2em\hbox{\rm R}}}}
\def\IZ{{\hbox{{\rm Z}\kern-.4em\hbox{\rm Z}}}}

\def\sIR{{\hbox{{\sevenrm I}\kern-.2em\hbox{\sevenrm R}}}}

%
%
\hyphenation{Min-kow-ski}

\rightline{HEP-UK-0005}
\rightline{August 1997}
\rightline{hep-th/9708083}

\vfill
%
%
\title{Finite N Heterotic Matrix Models 
and Discrete Light Cone Quantiztion}
\vfill

%
%

\author{Lubo\v s Motl\foot{motl@physics.rutgers.edu}}
\address{Department of Physics and Astronomy \break
 Rutgers University, Piscataway, NJ 08855-0849}
\vfill

\author{Leonard Susskind\foot{susskind@dormouse.stanford.edu}}
\address{Department of Physics \break Stanford University,
 Stanford, CA 94305-4060}
\vfill

%
%
The finite $N$ version of Matrix theory describes M-theory and superstrings
in so-called
discretized light cone quantization (DLCQ). Its role has been
explained for M-theory in 11 dimensions and for type IIA theory.
We show novelties which arise by generalizing the ideas to heterotic
strings. The states which arise in $O(N)$ theories with odd $N$ we interpret
as fields which are antiperiodic in the longitudinal direction.
Consistency of these ideas provides a new evidence for the conjecture
that finite $N$ models describe sectors with a given longitudinal momentum.

\vfill\endpage

%
%
\REF\bfss{T. Banks, W. Fischler, S. Shenker and L.
Susskind, hep-th/9610043.}
\REF\newcon{L. Susskind, hep-th/9704080.}
\REF\connew{L. Motl, Unpublished.}
\REF\DLCQone{T. Maskawa, K. Yamawaki, Prog. Theor. Phys. {\it 56} (1976)
270.}
\REF\DLCQtwo{A. Casher, Phys. Rev. {\it D14} (1976) 452.}
\REF\DLCQ{H.C. Pauli and S.J. Brodsky, Phys.Rev.{\bf D32} (1985) 1993.}
\REF\m{L. Motl, hep-th/9701025.}
\REF\bs{T. Banks and N. Seiberg, hep-th/9702187.}
\REF\dvv{R. Dijkgraaf, E. Verlinde and H. Verlinde, hep-th/9703030.}
\REF\bamo{T. Banks, L. Motl, hep-th/9703218.}
\REF\ulf{U. Danielsson, G. Ferretti, hep-th/9610082.}
\REF\kacheva{S. Kachru, E. Silverstein, hep-th/9612162.}
\REF\motlhet{L. Motl, hep-th/9612198.}
\REF\kimrey{N. Kim, S.-J. Rey, hep-th/9701139.}
\REF\horava{P. Horava, hep-th/9705055.}
\REF\wati{W.Taylor, hep-th/9611042.}
\REF\s{L. Susskind, hep-th/9611164.}
\REF\fhrs{W. Fischler, E. Halyo, R. Rajaraman and L. Susskind,
hep-th/9703102 }
\REF\ms{J. Maldacena and L. Susskind, Nucl. Phys. {\bf B475} (1996) 670,
hep-th/9604042.}
\REF\ss{S. Sethi and L. Susskind, hep-th/9702101.}

%
%

%
%
\chapter{Introduction}

According to ref's [\newcon,\connew]  Matrix Theory
is meaningful even for a finite size 
of the matrices and describe a particular
sector of states with a given value of longitudinal 
momentum in the so called
Discretized Light Cone Quantization (DLCQ). In DLCQ
[\DLCQone-\DLCQ], the lightlike
direction $x^-$ is compactified and thus the points with the 
same value of
$x^i,$ $i=1,\dots 9$ and $x^+$ (the light cone time)
and with $x^-$ differing by an integer multiple of $2\pi R$ are
identified.

Because of this compactification, the component $P^+$ of momentum
which is conjugate to $x^-$ must have quantized values $P^+=N/R$. Since
this component is conserved, the sectors of the Hilbert space with
a given value of $N$ are mutually decoupled. The physics of the
sector of M-theory with $P^+=N/R$ is described by the $U(N)$
supersymmetric quantum mechanics [\bfss]. The same is true not only
for M-theory, but also for its toroidal compactifications like
the IIA strings.

The low energy limit of M(atrix) theory usually allows a spacetime
interpretation and can be approximated by a quantum field theory.
The fields in such a field theory are periodic in $x^-$ with period
$R$. We will present a similar picture for heterotic strings where
some of the fields are antiperiodic in $x^-$.

\chapter{Short review of matrix strings}

After a compactification of $X^1$ to a circle, the compact coordinate is
represented by the covariant derivative with respect to a new spatial
dimension
$\sigma$, dual to $X^1$. The original
$0+1$ dimensional model [\bfss] describing M-theory in 11 dimensions
in DLCQ becomes a $1+1$ dimensional model describing M-theory on a circle
which is known to be equivalent to type IIA strings. Each string can have
longitudinal momentum $P^+=N/R$ and in the limit of zero coupling
(i.e. zero radius of $X^1$) it is approximated by matrices $X^i(\sigma)$
which are diagonalizable for each $\sigma$ and periodic up to a cyclic
permutation [\m] of length $N$. In [\bs] such solutions were explained
using the notion of the moduli space of the corresponding supersymmetric
Yang Mills theory.

Paper [\dvv] showed that the matrix strings reproduce the lowest order
interactions with a correct scaling law $\lambda\approx R^{3/2}$. It is
also expected that the matrix strings are able to give the higher order
contact interaction terms which are neccessary to fill the entire moduli
space of Riemann surfaces in the light cone gauge calculations.

The idea of supersymmetric Yang Mills theories living on the dual torus
can be continued up to $T^3$ compactification of M-theory. The central
charges are described by fluxes and base space momenta: the momentum is
$\int P^i$, the transverse membrane charges are $\int F^{ij}$ and the
longitudinal membrane charges $Z^{i-}$ are identified with momenta in the
base space. Note that in the case of matrix strings, $Z^{1-}$ corresponds
to strings' $L_0-\tilde L_0$.

\chapter{Short review of heterotic matrix models}

The heterotic matrix models were pioneered
in [\ulf] as the mechanics of D0-branes in type IA theory. In
[\kacheva] the new 16 vector fermionic degrees freedom corresponding
to strings joining D0-branes and D8-branes were found to be the source
of the $E_8$ symmetry. In [\motlhet] the model [\ulf] was exhibited
as an orbifold of the original model [\bfss]. This study showed that
the model describes M-theory with one boundary (and thus one $E_8$ group)
and contains also open membranes. Possible membrane topologies were
further investigated in [\kimrey] using the technology of area preserving
diffeomorphisms. In [\bamo] toroidal compactifications of heterotic
strings were first formulated via matrix models and the origin of various
sectors and GSO projections was explained.

Let us briefly discuss some technical matters. The original model [\bfss]
describes M-theory with $X^-$ compactified to a circle of radius $R$.
We can find a $Z_2$ symmetry [\motlhet] reversing sign of $X^1$
and multiplying spinors by $\gamma^1$.
But this reversing must be accompanied by transposition of all the
matrices which can be visualized by reversing of the orientation of
membrane.

Restricting the matrices containing the fundamental degrees of freedom
to be invariant under this operation, we restrict them to be real
(or pure imaginary) matrices and the original symmetry $U(N)$ is
reduced to $O(N)$: we have no condition for the determinant which
will be important for GSO projections and sectors -- and also we must
allow both even and odd values of $N$.
For consistency, we must also add 16 fermionic
vector degrees of freedom which are the source of $E_8$ symmetry.
These extra fields can be understood as a twisted sector of an
underlying theory generating the degrees of freedom of M(atrix) theory
and in lower dimensions their origin can be connected with
two-cycles of K3 manifolds shrunk to zero size.

We will normalize $R$ so that the period of $x^-$ is always $2\pi R$.
In the case of heterotic strings, the graviton appears first in
$N=2$ models, so we have the identity $P^+=N/2R$. The fields whose
states
appear already in $N=1$ models are antiperiodic in $x^-$ so their
$P^+$ is an odd multiple of $1/2R$. (Of course, we could keep the
condition $P^+=N/R$ but then the period in $x^-$ would be equal to
$\pi R$.) 
The $O(2N)$ model reduces to the $U(N)$ model far from
the domain wall; both models describe states with $P^+=N/R$.

Let us discuss the simplest (free) models $N=1$ describing states with
one quantum $P^+=1/2R$. For the [\bfss] model, we have $1\times 1$
matrices $P^i,X^i,\theta^i$ and the $P^-$ is given by $R\cdot P_i^2/2$.
The 16 fermions $\theta$ ensure the
$2^{16/2}=256$-fold degeneracy of the
state, reproducing 128 bosonic and 128 fermionic physical states
of supergravity.
For the [\ulf,\kacheva,\motlhet] model, only $X^2\dots X^9, 
P^2\dots P^9$ and
half $\theta$'s survive, but we have extra $\chi^1\dots \chi^{16}$. The
energy of massless states is given by
$$P^-={(P_i)^2\over 2P^+}=R\cdot P_i^2\eqn\hetenergy$$
but the degeneracy from
$\theta$'s is now only $2^{8/2}=16$ and there is an extra $256$-fold
degeneracy from $\chi$'s.
But not all 256 states are physical.
The gauge group is $O(1)$ which has only
one nontrivial element $T$, changing the sign of the 1-dimensional
vectors. It affects only the vectors $\chi$, so it anticommutes with
them. As a result, $T$ acts as the GSO operator, selecting only 128
from 256. So finally, the $O(1)$ model describes only $16.128$ states
which is the $SO(16)$ spinor part of the gauge supermultiplet.

For $O(2)$ symmetric system, we expect emergence of the graviton
supermultiplet and the 120 part ($SO(16)$ adjoint) of the gauge
supermultiplet. For higher values of $N$, the situation will be similar
to that of $N=1$ and $N=2$ for odd and even $N$, respectively.

Thus the shift in $x^-$ by $2\pi R$ is equivalent to the rotation
by $2\pi$ in the $SO(16)$ subgroups of the gauge groups $E_8$, under
which $120$ is even and $128$ odd. As was explained in [\bamo],
AP and PA sectors of the $E_8\times E_8$ heterotic string appears
for odd $N$ models while AA and PP sectors for even $N$ models.

We can also explain the nature of GSO projections for finite $N$.
There must be two GSO projections acting at the $\chi$ fields.
One GSO operator anticommutes with all the $\chi$ degrees of freedom.
This operator is given by the
global transformation given by the
$-1$ matrix from the $O(N)$ gauge group
and therefore even for finite $N$, only even states survive the condition
of invariance under the gauge group. In the case of the quantum mechanics
with one domain wall, this is the only projection we need.

But for the heterotic strings with two $E_8$ factors we need one extra
GSO operator anticommuting with half of $\chi$'s only. This projector is
not available for finite $N$ models but is only result of the large
$N$ limit, see [\bamo].

The situation here is similar to the level-matching conditions which
are also not available [\newcon] for finite $N$.  
Among the string states for
$N=1$, for instance, 
we find even the states which do not obey the conventional
level-matching conditions and the second GSO projection.
However, in the large  $N$ limit such unmatched states decouple.
The similarity between the second GSO projection and the level-matching
conditions is not an accident, of course. In AP and PA sectors,
fulfilling of level-matching conditions automatically implies the
GSO projection (whose GSO operator anticommutes with the
A half-integral fermions). In AA sector the level-matching condition
implies the GSO projection whose operator anticommutes with all the
$\chi$ fermions which is connected with the fact that for even $N$
where states of AA sector appear, $-1$ matrix belongs even to $SO(N)$.

\chapter{Broken $E_8$ symmetry}

Now we see why $E_8$ symmetry is not manifest in the heterotic matrix
models. Using the spacetime interpretation, the $E_8$ is not manifest
because it is broken by DLCQ procedures to $SO(16)$: going around
$x^-$ circle is equivalent to a nontrivial transformation in $E_8$
(which is however trivial for $SO(16)$ subgroup).

The situtation is similar to the nonmanifest Lorentz generators.
These generators are also not manifest because of the DLCQ machinery
which selects priviliged lightlike directions.

In both cases, the 128 generators of $E_8$ or the nontrivial
Lorentz generators mix the states with different values of $P^+$.
But also in both cases we can reconstruct the symmetry when the
period $R$ goes to infinity while keeping the $P^+$ fixed (since
we study states with particular finite values of $P^+$).
Because of the fixed $P^+$, large $R$ limit means large $N$ limit
and thus in the large $N$ limit we reconstruct whole $E_8$ symmetry.

For a possible $SO(32)$ symmetry, we could expect that the states
$(120,1)$ and $(1,120)$ of the $SO(16)\times SO(16)$ group appear
together with gravitons for even $N$ and the states $(16,16)$
for odd $N$. The interpretation will be completely analogous in
this case.

\chapter{Conclusion}

In this paper we described new evidence for the conjecture that
finite $N$ matrix models describe a sector with a given value
of longitudinal momentum in DLCQ: we explained how this works for
the heterotic matrix models. The new ingredient is the antiperiodicity
of some fields in the lightlike direction.

\refout
\end